\documentclass[11pt]{article}
\newsavebox{\PSLASH}
\sbox{\PSLASH}{$p$\hspace{-1.8mm}/}

\input{amssym}
\def\be{\begin{equation}}
\def\ee{\end{equation}}
\def\ba{\begin{eqnarray}}
\def\ea{\end{eqnarray}}
\begin{document}
\title{\large \bf Generalized Friedmann Equations for a Finite Thick Brane}
\author{S. Ghassemi$^{1}$\footnote{e-mail: Ghassemi@sharif.edu},
S. Khakshournia$^{2}$\footnote{e-mail: skhakshour@aeoi.org.ir},
and R. Mansouri$^{3}$\footnote{On sabbatical leave from Department
of Physics, Sharif University of Technology, Tehran. e-mail:
mansouri@hep.physics.mcgill.ca; mansouri@ipm.ir}  \\ \\
$^{1,3}$Department of Physics, Sharif University of Technology,\\
Tehran
11365-9161, Iran\\
$^{2}$Nuclear Research Center, Atomic Energy Organization of Iran,
Tehran, Iran\\
$^{3}$Institute for Studies in Physics and Mathematics(IPM), Tehran,\\
$^{3}$ Department of Physics, McGill University, 3600 University
Street, Montreal,\\ QC, H3A 2T8, Canada,}\maketitle
 \[\]
 \[ \]
\begin{abstract}
We present the generalized Friedmann equations describing the
cosmological evolution of a finite thick brane immeresed in a
five-dimensional Schwarzschild Anti-de Sitter spacetime. A linear
term in the density in addition to a quatratic one arises in the
Friedmann equation, leading to the standard cosmological evolution
at late times without introducing an ad hoc tension term for the
brane. The effective four-dimensional cosmological constant is
then uplifted similar to the KKLT effect and vanishes for a brane
thickness equal to the AdS curvature size, up to the third order
of the thickness. The four-dimensional gravitational constant is
then equal to the five-dimensional one divided by the AdS
curvature radius, similar to that derived by dimensional
compactification. An accelerating brane cosmology may emerge at
late times provided there is either a negative transverse pressure
component in the brane energy-momentum tensor or the effective
brane cosmological constant is positive.

\end{abstract}

\newpage

\section{Introduction}

An infinitesimally thin brane is a geometrical construction which
may reflect the characteristic features of the solitonic objects
found in string theory at the energy scales much smaller than the
energy scales related to the inverse thickness of the brane, or at
distance scales much bigger than the thickness of the brane.
However, once we are going to use the concept of brane
phenomenologically and apply it to cosmology we have to be careful
about the interplay of different scales inherent in a cosmological
model. For example, we have to be sure about the smallness of the
effect of thickness on the density or the cosmological parameter
before deciding to ignore it. This has never been shown explicitly.\\
Interest in walls and branes as solitonic localized matter
distributions, specially in higher dimensions, came from string
theory mainly because it provides a novel approach for resolving
the cosmological constant and the hierarchy problems \cite{Arkan}.
In this scenario, gravitation is localized on a brane reproducing
effectively four-dimensional gravity at large distances due to the
warp geometry of the spacetime \cite{RSII}. However, the history
of the interest in the localized matter distributions in the
context of gravity goes back to the early beginning of general
relativity. Recognizing the difficulty of handling thick walls
within relativity, already early authors considered the
idealization of a singular hypersurface as a thin wall and tried
to formulate its dynamics within general relativity \cite{SLD}.
Einstein and Strauss used implicitly the concept of a thick shell
to an embedded spherical star within a Friedmann-Robertson-Walker
universe \cite{Eins}. The new era of intense interests in thin
shells and walls began with the development of ideas related to
phase transitions in early universe and the formation of
topological defects. Again, mainly because of technical
difficulties, strings and domain walls were assumed to be
infinitesimally thin \cite{ViCv}.\\
Thereafter, interest in thin walls, or hypersurfaces of
discontinuity, received an impetus from the cosmology of early
universe. The formulation of dynamics of such singular
hypersurfaces was summed up in the modern terminology by Israel
\cite{israel}. Within the Sen-Lanczos-Israel (SLI) formalism, thin
shells are regarded as idealized zero thickness objects, with a
$\delta$-function singularity in their energy-momentum
and Einstein tensors.\\
In contrast to thin walls, thickness brings in new subtleties,
depending on how the thickness is defined and handled. Early
attempts to formulate thickness, being mainly motivated by the
outcome of the idea of late phase transition in cosmology
\cite{hill}, were concentrated on domain walls. Silveria
\cite{Sil} studied the dynamics of a spherical thick domain wall
by appropriately defining an average radius $<R>$, and then used
the well-known plane wall scalar field solution as the first
approximation to derive a formula relating $<\ddot{R}>$,
$<\dot{R}>$, and $<R>$ as the equation of motion for the thick
wall. Widrow \cite{Wid} used the Einstein-scalar equations for a
static thick domain wall with planar symmetry. He then took the
zero-thickness limit of his solution and showed that the
orthogonal components of the energy-momentum tensor would vanish
in that limit. Garfinkle and Gregory \cite{GG} presented a
modification of the Israel thin shell equations by using an
expansion of the coupled Einstein-scalar field equations
describing the thick gravitating wall in powers of the thickness
of the domain wall around the well-known solution of the
hyperbolic tangent kink for a $\lambda\phi^{4}$ wall and concluded
that the effect of thickness at first approximation was
effectively to reduce the energy density of the wall compared to
the thin case, leading to a faster collapse of a spherical wall in
vacuum. Others \cite{bar} applied the expansion in the wall action
and integrate it out perpendicular to the wall to show that the
effective action for a thick domain wall in vacuum apart, from the
usual Nambu term, consists of a contribution proportional
to the induced Ricci curvature scalar.\\
Study of thick branes in the string inspired context of cosmology
began almost simultaneously with the study of thin branes, using
different approaches. Although in brane cosmology the interest is
in local behavior of gravity and the brane, most of the authors
take a planar brane for granted \cite{Ell}. However, irrespective
of the spacetime dimension and the motivation of having a wall or
brane, as far as the geometry of the problem is concerned, most of
the papers are based on a regular solution of Einstein equations
on a manifold with specified asymptotic behavior representing a
localized scalar field \cite{Whole}. Some authors use a smoothing
or smearing mechanism to modify the Randall-Sundrum ansatz
\cite{Csak,GhKo}. Authors in \cite{GhKo} introduce a thickness to
the brane by smoothing out the warp factor of a thin brane world
to investigate the stability of a thick brane. In another approach
to derive generalized Friedmann equations, the four-dimensional
effective brane quantities are obtained by integrating the
corresponding five-dimensional ones along the extra-dimension over
the brane thickness \cite {Moun}. These cosmological equations
describing a brane of finite thickness interpolate between the
case of an infinitely thick brane corresponding to the familiar
Kaluza-Klein picture and the opposite limit of an infinitely thin
brane giving the unconventional Friedmann equation, where the
energy density enters quadratically. The latter case is then made
compatible with the conventional cosmology at late times by
introducing and fine tuning a negative cosmological constant in
the bulk and an intrinsic positive tension in the brane
\cite{whole1}. Recently, Navarro and Santiago \cite{Nav05}
considered a thick codimension 1 brane including a matter pressure
component along the extra dimension in the energy-momentum of the
brane. By integrating the 5D Einstein equations along the fifth
dimension, while neglecting the parallel derivatives of the metric
in comparison with the transverse ones, they write the equations
relating the values of the first derivatives of the metric at the
brane boundary with the integrated componentes of the brane
energy-momentum tensor. These, so called matching conditions are
then used to obtain the cosmological evolution of the brane  which
is of a non-standard type, leading to an accelerating universe for
special values of the model parameters.\\
A completely different approach based on the gluing of a thick
wall considered as a regular manifold to two different manifolds
on both sides of it was first suggested in \cite{khak02}. The idea
behind this suggestion is to understand the dynamics of a
localized matter distribution of any kind confined within two
metrically different spacetimes or matter phases. Such a matching
of three different manifolds is envisaged to have many diverse
applications in astrophysics, early universe, and string
cosmology. It enables one to have any topology and any spacetime
on each side of the thick wall or brane. The range of its
applications is from the dynamics of galaxy clusters and their
halos to branes in any spacetime dimension with any symmetry on
each side of it \cite{khak04}. By construction, such a matching is
regular and there is no singular surface whatsoever in this
formulation. Therefore Darmois junction conditions for the
extrinsic curvature tensors on the thick wall boundaries with the
two embedding spacetimes can be applied. \\
In this paper, we will use this formalism recently developed in
\cite{khos} for a finite thick wall and apply it, as an example,
to a thick brane embedded in a Schwarzschild Anti-de Sitter
(Sch-AdS) bulk to see the effect of thickness on the cosmology of
the brane. Although the dynamical equations can be written in an
exact form, to compare them with the standard cosmology we have to
make an expansion in terms of the brane thickness. It turns out
that the modified Friedmann equations are similar to the standard
one having a linear term in density, in contrast to the thin brane
cosmology, where the density term enters quadratically . \\
In section 2 we introduce our formalism to examine the
cosmological evolution of a thick brane embedded in the bulk
spacetime. Section 3 is devoted to the metric of the bulk and the
brane and the related quantities needed to be substituted in the
junction conditions. In section 4 we give the modified Friedmann
equations for the thick brane. Our conclusions are presented
in section 5.\\
\par
Throughout the paper we use $\Lambda$ for the five-dimensional
cosmological constant and $\kappa$ for its gravitational constant.
The two boundary limits of the thick brane are called $\Sigma_j$
with $j=1,2$. The core of the thick brane is denoted by
$\Sigma_0$. For any quantity $S$ let $S_{0}$ denote
$S|_{\Sigma_{0}}$.  Square bracket $[F]$ indicates the jump of any
quantity $F$ across $\Sigma_{j}$. Latin indices range over the
intrinsic coordinates of $\Sigma_{j}$ denoted by $\xi^{a}_{j}$,
and Greek indices over the coordinates of the 5-manifolds.

\section{Modelling the Thick Brane}

The technology of manipulating thin and thick localized matter
distributions or walls in general relativity in any dimension are
basically different. Thin walls can be treated in two different
but equivalent ways. Either one solves the Einstein equations in
$d+1$ dimension with a {\it distributional} energy-momentum tensor
which mimics an infinitesimally thin wall carrying some kind of
matter, dark or not dark, including radiation
, or one takes the known solutions of Einstein equations on either
side of the wall and glues them to the wall by applying the
boundary conditions at the wall location. The equivalence of these
two procedures is not trivial but has been shown rigorously for
the general case in \cite{MKH}. Boundary surfaces not carrying any
energy-momentum tensor can just be considered as a special case.
It should be noted that such an equivalence does not exist for
codimension 2 walls or defects, as it is also the case
for the strings in 4-dimensional spacetime.\\
The lack of such an equivalence in the case of thick walls or
localized matter distributions makes us differentiate between
different applications of the term thick walls or branes. Usually
a thick wall is considered to be a solution of Einstein equations
with a localized scalar field having a well-defined asymptotic
behavior. As mentioned in the introduction, we will continue to
use the term thick wall for such a solution of Einstein equations.
However, there is another case of interest with advantages in
diverse applications in astrophysics and string cosmology. Assume
a localized matter distribution to be considered as a solution of
Einstein equations on a specific manifold with well-defined
timelike boundaries. This localized or thick wall is then immersed
in a universe which could in principle consist of two different
solutions of Einstein equations on each side of the wall. The
combined manifold, consisting of three different solutions of
Einstein equations is again a solution of Einstein equations. The
localized wall may be infinite in the planar or cylindrical case
or compact in the spherical case. Therefore, in contrast to the
thin wall formalism where one glues two different manifolds along
a {\it singular} hypersurface, our definition of localized or
thick wall leads to the problem of gluing three different
manifolds along two {\it regular} hypersurfaces. \\
Let us now consider a thick codimension 1 brane immersed in a
5-dimensional bulk spacetime. Following the formalism introduced
in \cite{khos}, we take the thick wall with two boundaries
$\Sigma_1$ and $\Sigma_2$ dividing the overall spacetime $\cal M$
into three regions. Two regions $\cal M_{+}$ and $\cal M_{-}$ on
either side of the wall and the region ${\cal M}_{0}$ within the
wall itself. Treating the two surface boundaries $\Sigma_1$ and
$\Sigma_2$ separating the manifold ${\cal M}_{0}$ from two
distinct manifolds $\cal M_{+}$ and $\cal M_{-}$, respectively, as
nonsingular timelike hypersurfaces, we expect the intrinsic metric
$h_{ab}$ and extrinsic curvature tensor $K_{ab}$ of
$\Sigma_j\hspace{0.2cm}$ to be continuous across the corresponding
hypersurfaces . These requirements, the so-called Darmois
conditions, are formulated as
\begin{equation}\label{hmn}
[h_{ab}]_{\Sigma_j}=0\hspace {1cm} j=1,2,
\end{equation}
\begin{equation}\label{thick israel}
[K_{ab}]_{\Sigma_j}=0\hspace {1cm} j=1,2,
\end{equation}
where the square bracket denotes the jump of any quantity that is
discontinuous across $\Sigma_j$. To impose the Darmois conditions
on two surface boundaries of a given thick wall one needs to know
the metric in three distinct spacetimes $\cal M_{+}$, $\cal
M_{-}$ and ${\cal M}_{0}$ being jointed at $\Sigma_j$. While the
metrics in $\cal M_{+}$ and $\cal M_{-}$ are usually given in
advance, knowing the metric within the wall spacetime ${\cal M}_{0}$
requires a nontrivial work.\\
We assume the wall to have a proper thickness $2w$ in the fifth
dimension. We therefore have three different regions, two in the
bulk and one within the brane, to be joined together as three
different manifolds. Each boundary of the thick brane
($\Sigma_j\:\:,\: j=1,2$) glues the inside metric to a version of
the outside spacetime. From the two junction conditions for each
of the boundaries, to glue a slicing of the bulk to the spacetime
within the brane, we obtain:
\begin{equation}\label{kmn2}
K_{ab} \Bigl|^{+}_{\Sigma_{2}} -K_{ab}
\Bigl|^{-}_{\Sigma_{1}}+K_{ab} \Bigl|^{w}_{\Sigma_{1}}- K_{ab}
\Bigl|^{w}_{\Sigma_{2}}=0,
\end{equation}
where $+,-$ denote two slices of the outside spacetime and $w$
denotes the spacetime within the wall.\\
Now, we introduce a Gaussian normal coordinate system $(n,
\xi^{a}_{0})$ in the neighborhood of the core of the thick brane
denoted by $\Sigma_{0}$, where $\xi^{a}_{0}$ are the intrinsic
coordinates of $\Sigma_{0}$, and $n$ is the proper length along
the geodesics orthogonal to $\Sigma_{0}$ such that $n=0$
corresponds to $\Sigma_{0}$. Assuming that the brane thickness is
small in comparison with its curvature radius, we then expand the
extrinsic curvature tensor terms in the equation (\ref{kmn2}) in a
Taylor series around $\Sigma_{0}$ situated at $n=0$
\begin{equation}\label{Kexpan}
K_{ab}\Bigl|_{\Sigma_{i}}=K_{ab}
\Bigl|_{\Sigma_{0}}+\epsilon_{j}w\frac{\partial K_{ab}}{\partial
n} \Bigl|_{\Sigma_{0}} +O(w^2),
\end{equation}
where $\epsilon_{1}=-1$ and $\epsilon_{2}=+1$. The derivative of
the extrinsic curvature of the brane is related to the 5D
geometric quantities as follows:
\begin{eqnarray}\label{Kderiv}
\frac{\partial K_{ab}}{\partial
n}=K_{ad}K^d_b-R_{\mu\alpha\nu\sigma}n^{\alpha}
n^{\sigma}e^{\mu}_ae^{\nu}_b,
\end{eqnarray}
with $n^{\mu}$ being the normal vector field to the brane,
$R_{\mu\alpha\nu\sigma}$ the five-dimensional Riemann curvature
tensor and $e^{\mu}_{a}=\frac{\partial
x^{\mu}}{\partial\xi^{a}_{0}}$ are the four basis vectors tangent
to the brane. Substituting (\ref{Kderiv}) and (\ref{Kexpan}) into
(\ref{kmn2}) we arrive at the following equation being written on
the core of the thick brane
\begin{eqnarray}\label{junc}
K_{ab} \Bigl|^{+}_{\Sigma_{0}}-K_{ab} \Bigl|^{-}_{\Sigma_{0}}&+&
w\left((K_{ac}K^{c}_{b}-R_{\mu\sigma
\nu\lambda}e^{\mu}_{a}e^{\nu}_{b}n^{\sigma}n^{\lambda})
\Bigl|^{-}_{\Sigma_{0}} +(K_{ac}K^{c}_{b}-R_{\mu\sigma
\nu\lambda}e^{\mu}_{a}e^{\nu}_{b}n^{\sigma}n^{\lambda})
\Bigl|^{+}_{\Sigma_{0}}\right.\nonumber\\
&-&\left.2(K_{ac}K^{c}_{b}-R_{\mu\sigma\nu\lambda}
e^{\mu}_{a}e^{\nu}_{b}n^{\sigma}
n^{\lambda})\Bigl|^{w}_{\Sigma_{0}}\right)=0.
\end{eqnarray}
Having specified the metric of the bulk  and the metric within the
wall, all terms in the equation (\ref{junc}) are known and,
therefore, the dynamics of the thick brane is given.  In the limit
($w\rightarrow 0$) we should
also reproduce the familiar thin wall equations.\\

\section{Explicit calculations on the core}

Armed with Eq. (\ref{junc}) we now proceed to study the dynamics
of a thick brane of constant spacial curvature embedded in a
five-dimensional negative bulk cosmological constant. According to
the generalized Birkhoff's theorem the five-dimensional vacuum
cosmologically symmetric solution of Einstein's equations is
necessarily static and corresponds to Sch-AdS metric given by
\begin{equation}\label{metout}
ds^2=-f(r)dT^2+\frac{dr^2}{f(r)}+r^2d\Omega_{k}^{2},
\end{equation}
with
\begin{equation}\label{f}
f(r)=k-\frac{\Lambda}{6}r^2-\frac{C}{r^2},
\end{equation}
where $d\Omega_{k}^{2}$ is the metric of the 3D hypersurfaces
$\Sigma$ of constant curvature that is parameterized by $k=0, \pm
1$; $\Lambda$ is the bulk cosmological constant, and
constant $C$ is identified with the mass of a black hole located
at $r=0$. Since we need to know the spacetime of the thick brane
itself we here take the following ansatz for the metric of the
brane being written in a Gaussian normal coordinate system in the
vicinity of the core of the thick brane situated at $y=0$
\begin{equation}\label{metin}
ds^2=-n^2(t,y)dt^2+dy^2+a^2(t,y)d\Omega_{k}^{2},
\end{equation}
where $n^2(t,y)$ and $a^2(t,y)$ are some unknown functions to be
determined by solving the Einstein equations within the brane
with a suitable energy-momentum tensor, and $y$ is the normal
coordinate of the extra dimension. Compatibility with the
cosmological symmetries requires that the energy-momentum tensor
of the matter content in the brane takes the simple form
\begin{equation}\label{stress}
T^{\mu}_{\nu}=(-\rho,P_L,P_L,P_L,P_T),
\end{equation}
where the energy density $\rho$, the longitudinal pressure $P_L$,
and the transverse pressure $P_{T}$  are functions of $t$ and $y$.\\
Reading the metric (\ref{metin}) as
$ds^2=dy^2+\gamma_{ab}(x^{c},y)dx^{a}dx^{b}$, we now expand it in
a Taylor series in the vicinity of the core $\Sigma_{0}$ of the
thick brane placed at $y=0$ as follows:
\begin{equation}\label{taylor}
\gamma_{ab}(x^{c},y)=\gamma_{ab}(x^{c},0)+y\frac{\partial
\gamma_{ab}(x^{c},y)}{\partial
y}\Bigl|_{y=0}+\frac{y^2}{2}\frac{\partial^2
\gamma_{ab}(x^{c},y)}{\partial y^2}\Bigl|_{y=0}+O(y^3),
\end{equation}
where  $\gamma_{ab}(x^{c},0)$ is the metric on $\Sigma_{0}$, with
$x^{c}=(\tau, \chi, \theta, \varphi)$ the intrinsic coordinates of
$\Sigma_{0}$. But the derivatives in the expansion (\ref{taylor})
are given by
\begin{eqnarray}\label{Kd}
\frac{\partial \gamma_{ab}(x^{\sigma},y)}{\partial
y}&=&2K_{ab},\\
\frac{\partial^2 \gamma_{ab}(x^{\sigma},y)}{\partial
y^2}&=&2K_{ad}K^d_b-2R_{\mu\alpha\nu\sigma}n^{\alpha}
n^{\sigma}e^{\mu}_ae^{\nu}_b.
\end{eqnarray}
Let us write it more explicitly
\begin{equation}\label{exp1}
-n^2(t,y)=-n^2(t,0)+2yK_{\tau\tau}\Bigl|_{y=0}
+y^2(K^{\tau}_{\tau}K_{\tau\tau}-R_{tyty})\Bigl|_{y=0},
\end{equation}
\begin{equation}\label{exp2}
a^2(t,y)=a^2(t,0)+2yK_{\chi\chi}\Bigl|_{y=0}+y^2(K^\chi_\chi
K_{\chi\chi}-R_{\chi y\chi y})\Bigl|_{y=0}.
\end{equation}
Defining $a_{0}(\tau)=a(t(\tau), 0)$ and using the expansions
(\ref{exp1}) and (\ref{exp2}), we now write down the non-trivial
components of the full 5D Einstein's equations
$G_{\mu\nu}=\kappa^2T_{\mu\nu}$ at the location of the core of the
brane placed at $y=0$, with the metric (\ref{metin}) and the
energy-momentum tensor (\ref{stress}) as follows (see the Appendix
for the corresponding components of the Einstein tensor)
\begin{eqnarray}\label{EINSin}
ty&:&
-H_0K_{\tau\tau}+\frac{H_0}{a_{0}^2}K_{\chi\chi}-\frac{1}{a_{0}^2}
\dot{K}_{\chi\chi}=0,\\
yy&:&
\frac{K^2_{\chi\chi}}{a_0^4}-\frac{K_{\chi\chi}K_{\tau\tau}}{a_0^2}
-H_0^2-\frac{\ddot{a}_0}{a_0}
-\frac{k}{a_0^2}=\frac{\kappa^2}{3}P^{0}_T,\\
tt&:& H_0^2-\frac{1}{a_0^2}(B-k)
=\frac{\kappa^2}{3}\rho_{0}, \\
\chi\chi&:&\frac{2\ddot{a}_0}{a_0}+\frac{2K_{\tau\tau}K_{\chi\chi}}{a_0^2}
+\frac{K^2_{\chi\chi}}
{a_0^4}+K^2_{\tau\tau}-\frac{B}{a_0^2}+A=-\frac{\kappa^2}{3}
(\rho_{0}+3P^{0}_L),
\end{eqnarray}
where the dot stands for the derivative with respect to the proper
time $\tau$, $H_0=\frac{\dot{a}_{0}}{a_{0}}$,
$\rho_{0}=\rho(t,y=0)$, $P^{0}_{L}=P_{L}(t,y=0)$, and
$P^{0}_{T}=P_{T}(t,y=0)$. For the sake of brevity, we have defined
$A=K_{\tau\tau}K^{\tau}_{\tau}-R_{tyty}\Bigl|_{\Sigma_{0}}$,
$B=K^{\chi}_{\chi}K_{\chi\chi}- R_{\chi y\chi
y}\Bigl|_{\Sigma_{0}}$, and without loss of generality
$n(t, 0) = 1$.\\
Solving the $ty$ and $yy$ components of the Einstein equations for
$K_{\chi\chi}$ and $K_{\tau\tau}$, and performing the time
integration we obtain
\begin{equation}\label{Kchi}
K_{\chi\chi}\Bigl|^{w}_{\Sigma_{0}}=a_{0}\sqrt{\dot{a}^2_{0}
+\frac{2\kappa^2}{3}\frac{\tilde{P}_T}{a_{0}^2}+k+\frac{E}{a_{0}^2}},
\end{equation}
\begin{equation}\label{Ktt}
K_{\tau\tau}\Bigl|^{w}_{\Sigma_{0}}=\frac{\frac{2\kappa^2}{3}
\frac{\tilde{P_T}}{a_{0}^4} -\frac{\kappa^2}{3}P^{0}_T
-\frac{\ddot{a}_0}{a_0}+\frac{E}{a_{0}^4}}{\sqrt{H_0^2+\frac{2\kappa^2}{3}
\frac{\tilde{P_T}}{a_{0}^4}+\frac{k}{a_{0}^2}+\frac{E}{a_{0}^4}}},
\end{equation}
where $E>0$ is an integration constant, and
\begin{equation}\label{tildeP}
\tilde{P_T}\equiv\int_{0}^{\tau} P^{0}_T\:a^4_0\:H_0d\tau.
\end{equation}
From $tt$ component of the Einstein equations one can quickly read
\begin{equation}\label{kxxin} B=(K^{\chi}_{\chi}K_{\chi\chi}- R_{\chi y\chi
y})\Bigl|^w_{\Sigma_{0}}=H_0^2a_0^2+k -\frac{\kappa^2}{3}\rho
a_0^2.
\end{equation}
Furthermore,  substituting the expressions  (\ref{Kchi}) and
(\ref{Ktt}) into $\chi\chi$ component of the Einstein equations
yields the following expression for $A$
\begin{equation}\label{Aaccelerate}
A=K^{\tau}_{\tau}K_{\tau\tau}-
R_{tyty}\Bigl|^{w}_{\Sigma_{0}}=-K_{\tau\tau}^{2}\Bigl|^{w}_{\Sigma_{0}}
-\frac{2\kappa^2}{3}\left(\rho_{0}+\frac{3}{2}P^{0}_{L}-P^{0}_{T}
+\frac{3}{a_{0}^4}\tilde{P_T}\right)-\frac{3E}{a_{0}^4}.
\end{equation}
Returning now to the Sch-AdS bulk spacetime (\ref{metout}),  we
note that the corresponding four velocity $u^{\mu}$ and the normal
vector $n^{\mu}$ being evaluated on the thick brane's core
$\Sigma_{0}$ are, respectively
\begin{equation}\label{unout}
u^{\mu}\Bigl|^{\pm}_{\Sigma_{0}}=(\dot{T},\dot{a},0,0,0)
\Bigl|_{\Sigma_{0}},\hspace{1cm}
n^{\mu}\Bigl|^{\pm}_{\Sigma_{0}}=\varepsilon_{\pm}\left
(f^{-1}\dot{a},f\dot{T},0,0,0\right)\Bigl|_{\Sigma_{0}},
\end{equation}
where the sign function $\varepsilon= \pm{1}$ takes care of the
different patches of the Sch-AdS spacetime which might be glued to
the brane, and
$\dot{T}|_{\Sigma_{0}}=\frac{\sqrt{f_{0}+\dot{a}_{0}^2}}{f_{0}}$.
Subsequently, the relevant components of the extrinsic curvature
tensor on $\Sigma_{0}$ computed from the Sch-AdS metric
(\ref{metout}) are
\begin{eqnarray}\label{Kout1}
K_{\chi\chi}\Bigl|^{\pm}_{\Sigma_{0}}&=&\varepsilon_{\pm}
a_{0}\sqrt{f_{0}+\dot{a}_{0}^2},
\end{eqnarray}
\begin{eqnarray}\label{Kout2}
K_{\tau\tau}\Bigl|^{\pm}_{\Sigma_{0}}&=&-\frac{\varepsilon_{\pm}}{\sqrt{f_{0}
+{\dot{a}_{0}}^2}}\left(\ddot{a}_{0}
-\frac{\Lambda}{6}a_{0}+\frac{C}{a_{0}^3}\right),
\end{eqnarray}
where $f_{0} = f(r=a_{0}(\tau))$. The assumption of the $Z_2$
symmetry, with $y=0$ as a fixed point, leads to
$K_{ab}\Bigl|^{+}_{\Sigma_{0}}=-K_{ab}\Bigl|^{-}_{\Sigma_{0}}$,
implying that for matching of two interior patches of the Sch-AdS
spacetime one has to choose $ \varepsilon _{+}=-1$ and $
\varepsilon _{-}=+1$. This choice of the patches of the  Sch-AdS
metric is necessary to have the $Z_2$ symmetry which makes the
problem simpler.\\
Moreover, from the metric (\ref{metout}), the nonzero components
of the Riemanian curvature tensor on $\Sigma_{0}$ are calculated
to be
\begin{eqnarray}\label{Riemanout}
R_{\chi r\chi
r}\Bigl|^{w}_{\Sigma_{0}}=\frac{1}{f_{0}}\left(\frac{\Lambda
a_{0}^2}{6} -\frac{C}{a_{0}^2}\right),\hspace{0.1cm}R_{\chi T\chi
T}\Bigl|^{w}_{\Sigma_{0}}=f_{0}\left(\frac{-\Lambda a_{0}^2}{6}
+\frac{C}{a_{0}^2}\right),\hspace{0.1cm}
R_{TrTr}\Bigl|^{w}_{\Sigma_{0}}=\frac{\Lambda}{6}-\frac{3C}{a_{0}^4}.
\end{eqnarray}
We are now ready to write down explicitly the dynamical equations
of the brane using the master equation (\ref{junc}).\\

\section{Generalized Friedmann Equations}

Substituting Eqs. (\ref{kxxin}), (\ref{unout}), (\ref{Kout1}), and
(\ref{Riemanout}) into the $\chi\chi$ component of the equation
(\ref{junc}) we obtain
\begin{equation}\label{f1}
\sqrt{f_{0}+\dot{a}_{0}^2}=\frac{w}{a_{0}}\left(-\frac{\Lambda}{3}a_{0}^2
+\frac{\kappa^2}{3}\rho_{0}a_{0}^2\right).
\end{equation}
Taking the square of Eq. (\ref{f1}), substituting the expression
(\ref{f}) and rearranging, we arrive at the following equation
\begin{equation}\label{f3}
H_{0}^2+\frac{k}{a_{0}^2} =
\frac{2\kappa^2w^2(-\Lambda)}{9}\rho_{0}
+\frac{\kappa^4w^2}{9}\rho_{0}^2+\left(\frac{\Lambda}{6}
+\frac{w^2\Lambda^2}{9}\right)+\frac{C}{a_{0}^4}.
\end{equation}
Assuming the brane energy density profile as a Taylor series
around $y=0$
\begin{equation}\label{profrou}
\rho(t,y)=\rho_{0}+y\frac{\partial\rho}{\partial
y}\Bigl|_{y=0}+\frac{1}{2}y^{2}\frac{\partial^{2}\rho}{\partial
y^{2}}\Bigl|_{y=0}+O(y^3),
\end{equation}
we conveniently define an effective four-dimensional energy
density $\varrho$ associated to the five-dimensional energy
density $\rho$ as
\begin{equation}\label{rho}
\varrho=\int_{-w}^{w} \rho dy\simeq 2w\rho_{0}+O(w^2).
\end{equation}
We then identify
\begin{equation}\label{effect}
\frac{\Lambda_4}{3}=\frac{\Lambda}{6}
+\frac{w^2\Lambda^2}{9},
\end{equation}
\begin{equation}\label{GN}
8\pi G = \frac{\kappa^2w(-\Lambda)}{3}.
\end{equation}
Putting all this together, Eq. (\ref{f3}) turns into the following
form
\begin{equation}\label{f4}
H_{0}^2+\frac{k}{a_{0}^2} = \frac{8\pi
G}{3}\varrho+\frac{\kappa^4}{36}\varrho^2
+\frac{\Lambda_4}{3}+\frac{C}{a_{0}^4}.
\end{equation}
This equation is our main result. As we see, there is a linear in
addition to a quadratic term in the matter density, due to the
non-vanishing of the thickness $w$, which is a novel effect. There
is no need of introducing an ad hoc tension for the brane, and
splitting it from the matter density on the brane. According to
this equation the cosmological expansion undergoes a transition
from a high energy regime $\kappa^2\varrho\gg w(-\Lambda)$, where
the dominated $\varrho^2$ term yields the unconventional
cosmological expansion, into a low energy regime
$\kappa^2\varrho\ll w(-\Lambda)$ where the brane observers recover
the standard cosmology described by the usual Friedmann equation.
The compatibility with the Big Bang Nucleosynthesis (BBN) puts an
essential constraint on the parameters of the model, so that to
preserve the predictions of standard cosmology, the high energy
regime, where the $\varrho^2$ term is significant, must occur
before BBN era. From Eqs. (\ref{GN}) and (\ref{f4}) this implies
that $\kappa^{-2}w|\Lambda|\geq (1 MeV)^4$, yielding the
constraint $M\geq 10^4 GeV$, for the fundamental mass scale
defined by $\kappa^2=M^{-3}$, the same result as obtained in the
thin brane case.\\
Taking the thickness of the brane equal to the curvature size of
the AdS defined as $\Lambda = \frac{-6}{l^2}$, we obtain from
(\ref{effect}) exactly $\Lambda_4 = 0$. This means that the
effective 4-dimensional cosmological constant induced on the brane
vanishes up to the third order in the thickness. A residual term
proportional to the third order may still remain. This may be a
hint to the solution of the cosmological constant problem!\\
The effective brane cosmological constant has an interesting
behavior too. According to (\ref{effect}), it is up-lifted
relative to its value in the AdS bulk. This effect is similar to
the KKLT up-lifting of the AdS minimum by inclusion of $\overline
{D3}$ branes in the warped geometry put forward in \cite{kklt},
derived from purely geometrical considerations. If there is any
deep connection to the KKLT
uplifting effect remains to be seen. \\
Assuming again $2w = l$, from (\ref{GN}), the 4-dimensional
gravitational constant then becomes  $8\pi G =
\frac{\kappa^2}{l}$. This is exactly the value derived by the
dimensional compactification of the fifth dimension. Another
consequence worth mentioning is the proportionality of the
four-dimensional Newton's constant given by (\ref{GN}) to the
brane thickness. Assuming a time dependent brane thickness, this
induces a time evolution for the Newton's constant. On the
cosmological scale this is experimentally constrained \cite{Chib},
imposing tight restrictions on the time dependence
of the brane thickness.\\
The thin brane limit of our thick brane is also easily derived. In
this limit the Eq. (\ref{f4}) reduces to the unconventional
Friedmann equation of thin brane cosmology \cite{whole1}
\begin{equation}\label{ft}
H_{0}^2+\frac{k}{a_{0}^2}=\frac{\kappa^4}{36}\varrho^2
+\frac{\Lambda}{6}+\frac{C}{a_{0}^4},
\end{equation}
We may also look at the acceleration equation and its thin brane
limit. Take the expressions (\ref{Aaccelerate}), (\ref{unout}),
(\ref{Kout2}), and (\ref{Riemanout}) to write down the $\tau\tau$
component of Eq. (\ref{junc}) explicitly. We then end up with the
following equation:
\begin{eqnarray}\label{acc}
\frac{\frac{\ddot{a}_{0}}{a_{0}}-\frac{\Lambda}{6}
+\frac{C}{a_{0}^4}}{\sqrt{f_{0}+
\dot{a}_{0}^2}}&=&\frac{-w}{a_{0}}\left(
\frac{2\kappa^2}{3}\left(\rho_{0}
+\frac{3}{2}P^{0}_{L}-P^{0}_{T}+\frac{3}{a_{0}^4}\tilde{P_T}\right)
+\frac{\Lambda}{6}+\frac{3E}{a_{0}^4} +\frac{3C}{a_{0}^4}\right),
\end{eqnarray}
in deriving (\ref{acc}), we see from Eqs. (\ref{thick israel}) and
(\ref{Kexpan}) that $K_{\tau\tau}
\Bigl|^{+}_{\Sigma_{0}}-K_{\tau\tau}
\Bigl|^{w}_{\Sigma_{0}}=O(w)$, and can therefore be neglected.\\
Now, defining the four-dimensional effective quantities associated to
the five-dimensional longitudinal and transverse pressures $P_{L}$
and $P_{T}$ in the form
\begin{equation}\label{plfourdim}
 p_{L}=\int_{-w}^{+w}P_{L}\:dy\simeq 2wP^0_{L}+O(w^2),
 \end{equation}
\begin{equation}\label{ptfourdim}
 p_{T}=\int_{-w}^{+w}P_{T}\:dy\simeq 2wP^0_{T}+O(w^2),
 \end{equation}
we realize that in the zero thickness limit ($w\rightarrow 0$),
Eq. (\ref{acc}) reduces to the Raychaudhuri equation for a thin
brane \cite{Bin99}:
\begin{equation}
\frac{\ddot{a}_{0}}{a_{0}}+\frac{\dot{a}_{0}^2}{a_{0}^2}
+\frac{k}{a_{0}^2}=-\frac{\kappa^4}{36}\varrho(\varrho
+3p_{L})+\frac{\Lambda}{3},
\end{equation}
where we have used the thin brane equation (\ref{ft}) and the fact
that the  profile for the transverse pressure $P_{T}$ will not
blow up in the thin brane limit. Therefore, we get the familiar
thin brane limit from our thick brane model, as would be
expected.\\
Note now that the time component of the covariant derivative of
the brane energy-momentum tensor (\ref{stress}), using the metric
(\ref{metin}), leads to the familiar energy conservation condition
on the core of the thick brane:
\begin{equation}\label{conser}
\dot{\rho_{0}}+3H_{0}(\rho_{0}+P^0_{L})=0.
\end{equation}
Let us assume the arbitrary effective equations of state of the
form
\begin{equation}\label{state}
 p_{L}=\omega_{L}\varrho,\hspace{1cm}p_{T}=\omega_{T}\varrho,
\end{equation}
with constants $\omega_{L}$ and  $\omega_{T}$. The conservation
equation (\ref{conser}) can then be integrated with the result as
usual
\begin{equation}\label{conser2}
\rho_{0}=\rho_{i}a_{0}^{-3(1+\omega_{L})},
\end{equation}
where $\rho_{i}$ is a constant. Then $\tilde{P_T}$, according to the
definition (\ref{tildeP}), can be computed as
\begin{equation}\label{tildePevol}
\tilde{P_T}=\frac{\rho_{i}\omega_{T}}{1-3\omega_{L}}a_{0}^{1-3\omega_{L}}.
\end{equation}

Of great cosmological interest is the possibility of a late time
accelerated expansion on the brane. To investigate
this possibility, we take a closer look at the
generalized acceleration equation (\ref{acc}). Inserting Eqs.
(\ref{state}), (\ref{conser2}), and (\ref{tildePevol}), Eq. (\ref{acc})
can be recast as
\begin{eqnarray}\label{acc2}
\frac{\ddot{a}_{0}}{a_{0}}&=&
-\frac{\sqrt{f_{0}+\dot{a}_{0}^2}}{a_{0}}\frac{\kappa^2\varrho}{3}
\left(1+\frac{3\omega_{L}}{2}-\omega_{T}
+\frac{3\omega_{T}}{1-3\omega_{L}}\right)
+\frac{1}{6}\left(\Lambda_{4}+\frac{\Lambda}{2}\right)\nonumber\\
&+&\frac{\kappa^2\varrho w}{6}\left(\frac{-\Lambda}{6}\right)
-\frac{3w\sqrt{f_{0}+\dot{a}_{0}^2}}{a_{0}}\left(\frac{E}{a_{0}^4}
+\frac{C}{a_{0}^4}\right)-\frac{C}{a_{0}^4},
\end{eqnarray}
where we have used Eqs. (\ref{f1}) and (\ref{effect}).  Note that
at low energies, i.e. at late times, the two last terms on the
right hand side of Eq. (\ref{acc2}) redshift quickly and one can
then neglect them. Hence, this equation tells us that for the
acceleration term on the left hand side of Eq. (\ref{acc2}) to be
positive one or both of the following conditions must be satisfied
\begin{equation}\label{cond1}
1+\frac{3\omega_{L}}{2}-\omega_{T}
+\frac{3\omega_{T}}{1-3\omega_{L}}<0.
\end{equation}
\begin{equation}\label{cond2}
\Lambda_4 > \frac{-\Lambda}{2}
\end{equation}
In particular, in the case of $\omega_{L}=0$ for dust matter, the
constraint (\ref{cond1}) immediately reduces to
$\omega_{T}<-\frac{1}{2}$. Consequently, we see that an
accelerated cosmological expansion on the core of the thick brane
is possible if one includes the matter having a negative pressure
along the extra dimension in the brane energy-momentum tensor, or
if the effective 4-dimensional cosmological constant is positive,
according to the condition (\ref{cond2}) (see also \cite{Nav05}).

\section{Conclusion}

We have obtained a general equation (\ref{junc}) to study the
dynamics of codimension one brane of finite thickness immersed in
an arbitrary bulk spacetime. This was obtained in a general
setting by imposing the Darmois junction conditions on the brane
boundaries with the two embedding spacetimes and then using an
expansion scheme for the extrinsic curvature tensor at the brane
boundaries in terms of the proper thickness of the brane. Our
formalism is valid for any brane whose thickness is small compared
to its curvature radius. Using this approach we gave the
generalized Friedmann equations written for expansion of the
extrinsic curvatures up to the first order of the brane proper
thickness governing the cosmological evolution of the core of a
thick brane embedded in a five-dimensional Schwarzschild Anti-de
Sitter spacetime with a $Z_{2}$ symmetry. The derived equations
for the thick brane have the well-known limit of the thin brane
equations.\\
There are some novel effects for this finite thick brane
cosmology. First, the generelized Friedmann equation shows a
linear in addition to a quadratic term in the density. Therefore,
the late time behavior is the same as the standard cosmology
without introducing an ad hoc brane tension into the
energy-momentum tensor of the brane. Second, the effective induced
4-dimensional cosmological constant of the brane is increased
similar to the KKLT uplifting of the AdS minimum. It turns out
that this 4-dimensional cosmological constant vanishes for a
thickness equal to the AdS curvature size, up to the third power
of the thickness. The 4-dimensional Newton's gravitational
constant is then equal to the 5-dimensional one divided by the AdS
length, which is similar to the result  derived through the
dimensional compactification. According to the equation
(\ref{acc2}), the universe filled with dust matter, will be
accelerating at late times, if the pressure along the extra
dimension in the brane energy-momentum tensor is negative, or if
the effective cosmological constant satisfies the condition
(\ref{cond2}) .

\section{Acknowledgments}
RM would like to thank members of the cosmology group at McGill
University, specially Robert Brandenberger and Keshav Dasgupta for
helpful discussions. and McGill Physics department and Robert
Brandenberger for the hospitality.

\section{Appendix}

The non-trivial components of the five-dimensional Einstein tensor
$G_{\mu\nu}$ for the metric (\ref{metin}) are computed as
\begin{eqnarray}
{ G}_{00} &=& 3\frac{\dot{a}^2}{a}  - 3{n^2} \left({a''\over
a}+{{a'}^2\over a^2}\right)
 +3 k \frac{n^2}{a^2},
\label{ein00} \\
 { G}_{ij} &=&
a^2 \gamma_{ij}\left(2\frac{a^{\prime \prime}}{a} +\frac{n^{\prime
\prime}}{n}+ {{a'}^2\over a^2} +2{a^\prime n^\prime\over n}\right)\nonumber \\
& &+\frac{a^2}{n^2} \gamma_{ij} \left( -2{\ddot a\over a}-
{\dot{a}^2\over a^2}+2{\dot{a}\dot{n}\over an} \right) -k
\gamma_{ij},
\label{einij} \\
{G}_{0y} &=&  3\left({n^\prime\over n} {\frac{\dot{a}}{a}}  -
\frac{\dot{a}^{\prime}}{a} \right),
\label{ein05} \\
{G}_{yy} &=& 3\left( {{a'}^2\over a^2} +{a^\prime n^\prime\over
an} \right) - \frac{3}{n^2} \left({\ddot a\over
a}+{\frac{\dot{a}^2}{a^2}} -\frac{\dot{a}}{a} \frac{\dot{n}}{n}
\right) - 3 \frac{k}{a^2}, \label{ein55}
\end{eqnarray}
where a dot stands for a derivative with respect to $t$ and a
prime a derivative with respect to $y$.

\end{document}